\documentclass[prl,amsfonts,twocolumn,superscriptaddress,showpacs]{revtex4}

\usepackage{epsf}
\usepackage{amsmath}

\begin{document}

\def\be{\begin{equation}}
\def\ee{\end{equation}}
\def \bea#1\eea {\begin{eqnarray}#1\end{eqnarray}}
\def\Tr{\mathop{\rm Tr}}
\def\br{{\bf r}}
\def\bq{{\bf q}}
\def\vp{\varphi}
\newcommand{\corr}[1]{\langle #1\rangle}
\newcommand{\sign}{\mathop{\rm sign}}
\newcommand{\eps}{\varepsilon}
\newcommand{\const}{\text{const}}
\def\cD{{\cal D}}
\def\cC{{\cal C}}

\title{Dephasing in disordered metals with superconductive grains}

\author{M. A. Skvortsov}
\email{skvor@itp.ac.ru}
\affiliation{L. D. Landau Institute for Theoretical Physics, Moscow 119334, Russia}

\author{A. I. Larkin}
\affiliation{L. D. Landau Institute for Theoretical Physics, Moscow 119334, Russia}
\affiliation{Department of Physics, University of Minnesota, Minneapolis, MN 55455, USA}

\author{M. V. Feigel'man}
\affiliation{L. D. Landau Institute for Theoretical Physics, Moscow 119334, Russia}

\begin{abstract}
Temperature dependence of electron dephasing time $\tau_{\varphi}(T)$ is
calculated for a disordered metal with small concentration of
superconductive grains. Above the macroscopic superconducting transition
line, when electrons in the metal are normal, Andreev reflection from the
grains leads to a nearly temperature-independent contribution to the
dephasing rate. In a broad temperature range $\tau_{\varphi}^{-1}(T)$
strongly exceeds the prediction of the classical theory of dephasing in
normal disordered conductors, whereas magnetoresistance is dominated (in
two dimensions) by the Maki-Tompson correction and is positive.
\end{abstract}

\date{\today}

\pacs{74.40.+k, 72.15.Rn, 74.50.+r}

\maketitle

\section{Introduction}

During last few years, a number of experimental data on
electron transport in disordered metal films and wires were shown to be in
disagreement with the standard theory~\cite{AAK} of
electron dephasing in normal conductors.
Namely, at sufficiently low
temperatures $T \leq T_1$ the dephasing rate $\tau_{\varphi}^{-1}(T)$ was
systematically found to deviate from the power-law dependence~\cite{AAK}:
\be
\label{tauAAK}
  \frac{1}{\tau_\varphi^{(0)}(T)}
  =
  \begin{cases}
  \displaystyle
  \sim (T/\hbar)^{3/2}\tau^{1/2}/(k_Fl)^2 ,
    & \text{3D case} ,
  \\[5pt]
  \displaystyle
  (T/2\pi\hbar g) \ln (\pi g) , \quad
    & \text{2D case} ,
  \end{cases}
\ee
with a tendency to apparent saturation in the $T\to 0$ limit
($g = \hbar/e^2R_{\Box} \gg 1$ is dimensionless conductance of the film).
Since no dephasing rate may exist at strictly zero
temperature~\cite{Imry}, such a behavior indicates a presence of some
additional temperature scale(s) $T_0$
(which may occur to be extremely low), so that in the range $T_0 \leq T
\leq T_1$ the main contribution to $\tau_{\varphi}^{-1}(T)$ comes from some
new mechanism, different from the universal Nyquist noise considered in
Ref.~\cite{AAK}. Among other suggestions (the presence of localized two-level
systems~\cite{TLS1,TLS2}, nonequilibrium noise~\cite{AltKrav}, etc.)\/
there were some speculations on a possible role of electron-electron
interactions in $\tau_{\varphi}(T)$ ``saturation''.
Recent development~\cite{Aleiner} of the
theory~\cite{AAK} have proved that {\em perturbative}\/ account of
electron-electron interactions does not lead to considerable
corrections to Eq.~(\ref{tauAAK}).

In this paper we show that
electron-electron interaction considered {\em nonperturbatively}\/ can
indeed be responsible for strong deviation of dephasing rate from the
standard predictions. Namely, we consider a system of small
superconductive islands (of characteristic size $a$) situated in either
bulk disordered metal matrix (3D case) or on the thin metal film (2D).
The role of interaction here is to establish superconductivity in the
islands, which is a nonperturbative effect.
Such a system can exhibit~\cite{FL98} a macroscopic
superconducting transition mediated by the proximity Josephson
coupling between the islands~\cite{ALO68},
with the transition temperature $T_c(n_i)$
depending on the concentration of the islands $n_i$.
Above this transition electrons in the metal are normal,
but Andreev reflection of them from the superconducting islands
leads to an additional contribution to the dephasing rate:
\be
  1/\tau_\varphi(T)
  =
  1/\tau_\varphi^{(0)}(T)
  +
  1/\tau_\varphi^{\text{A}}(T) .
\label{t+t}
\ee

Enhancement of dephasing rate due to superconductive fluctuations
in {\em homogeneous}\/ systems
was considered previously both experimentally~\cite{Gordon84} and
theoretically~\cite{Patton}. Far above $T_c$,
the dephasing rate due to interaction in the Cooper channel
is comparable to the dephasing rate $\hbar/\tau_\varphi^{(0)}(T)$
due to the Coulomb interaction, being additionally suppressed
as $1/\ln^2(T/T_c)$.
Peculiarity of our result is that the superconductive
contribution to the dephasing rate in {\em inhomogeneous}\/
systems can be the dominant one
in a broad range of temperatures above $T_c(n_i)$.

To simplify calculations, we consider the model system~\cite{FL98} where
superconducting (SC) islands are connected
to the metal matrix via tunnel barriers with
normal-state tunnel conductances $G_T$ (measured in units of $e^2/\hbar$),
which determine inter-islands resistance in
the normal state. We are interested in the temperature range much
below the critical temperature $T_{c0}$ of islands, when charge transport
between them and the metal occurs due to Andreev reflection processes.
We assume large Andreev conductance, $G_A \gg 1$, thus
Coulomb blocking of Andreev transport is suppressed.
For small concentration of the islands, $n_i<n_c\sim\exp(-\pi G_A/4)$,
quantum fluctuations destroy macroscopic superconductive
coherence through the whole system even at $T=0$~\cite{FL98,FLS01}.
In the opposite limit, $n_i \gg n_c$, the thermally driven
superconductor--metal transition takes place
at $T_c(n_i)\sim \hbar Dn_i^{2/d}$, where $D$ is the
diffusion coefficient and $d$ is the dimensionality of space.

\begin{figure}
\epsfxsize=75mm
\centerline{\epsfbox{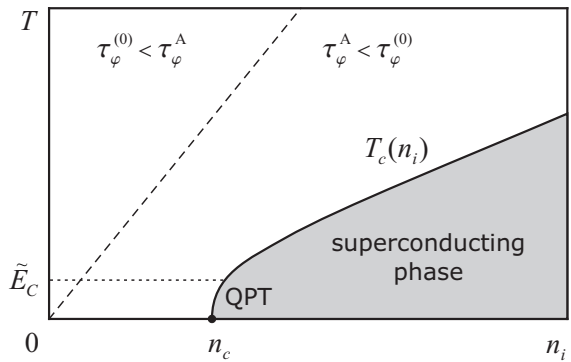}}
\caption{Schematic $(n_i,T)$ phase diagram of a metal with
superconducting grains. The dephasing time $\tau_\vp^A$
due to Andreev reflection is shorter than $\tau_\vp^{(0)}$
in a broad range above $T_c(n_i)$.}
\label{F:Tn}
\end{figure}

Here we focus on the temperature scale $T \gg T_c(n_i)$,
where macroscopic superconductivity is destroyed by thermal fluctuations,
and the phases $\varphi_j$ of superconductive order parameters on different
islands fluctuate strongly and are uncorrelated with each other.
Our main result is the expression for the dephasing rate due to
the processes of Andreev reflection from the SC islands:
\be
  \frac1{\tau_\varphi^{\text{A}}(T)}
  =
  \begin{cases}
  \displaystyle
    \frac{n_i}{4\nu\hbar}
    \left[ G_A - \frac4{\pi}\ln\frac{G_AE_C}{2\pi^2 T} \right] ,
      & \text{3D case} ,
  \\[10pt]
  \displaystyle
    \frac{n_i}{4\nu\hbar} \, G_A(T) ,
      & \text{2D case} ,
  \end{cases}
\label{result}
\ee
where
\be
  G_A(\omega) = \frac{G_T^2}{G_D(\omega)}
\label{GA}
\ee
is the (frequency-dependent) Andreev conductance of the island
in the lowest tunneling approximation~\cite{Nazarov94},
with $G_D^{-1} = (e^2/\hbar) (4\pi \sigma a)^{-1}$ for 3D spherical islands
of radii $a$, and $G_D^{-1}(\omega) = (4\pi g)^{-1} \ln(D/a^2\omega)$,
for 2D islands of radii $a$.
Here $\sigma$ is the 3D conductivity of the metal matrix,
$g=2\hbar \nu D \gg 1$ is the dimensionless film conductance per square,
$E_C= 2e^2/C$ is the bare charging energy of an island
and $\nu$ is the metal density of states per one spin projection.

Equation (\ref{result}) is valid for $T\gg \max(T_c(n_i),\tilde{E}_C)$,
where $\tilde{E}_C \propto E_C e^{-\pi G_A/4} $ is the renormalized
charging energy (see below).
In this temperature range the dephasing rate (\ref{result})
is nearly temperature
independent, thus exceeding the result~(\ref{tauAAK})
for sufficiently small $T < T_*(n_i) \sim G_A^{2/d}(T) \, T_c(n_i)$.
Therefore, the window where Andreev reflection off the islands
is the dominating dephasing mechanism is wide provided that $G_A(T)\gg1$.

In three dimensions we can also study the limit $T\ll \tilde E_C$
available at $n_i\ll n_c$, where macroscopic superconductivity
never occurs due to quantum fluctuations.
Here, the dephasing rate $1/\tau_\vp^A \propto (T/\tilde E_C)$
[see Eq.~(\ref{lowT})]
vanishes at $T\to0$ in accordance with
the general statement of Ref.~\cite{Imry}.

Below we provide brief derivation of the
result~(\ref{result}) and then discuss its physical origin and
implications for observable $\tau_\varphi (T)$ in 3D and 2D systems.

\section{Description of the formalism}

We start from the action functional $S =  S_D +  S_T$
for the disordered metal ($S_D$) and tunnel junctions with
SC islands ($ S_T$), written in the replica form of the
imaginary-time $\sigma$-model~\cite{Finkel,Oreg,FKLS}:
\bea
  S_D = \frac{\pi\nu}{8} \Tr \left[
    D (\nabla Q)^2 - 4\tau_3EQ
  \right] ,
\label{SD}
\\
  S_T = - \frac{\pi \gamma}{8} \sum_j \int dA_j \Tr Q(\br_j) Q_{Sj} .
\label{ST}
\eea
Integration in Eq.~(\ref{ST}) goes over
the  contact areas $A_j$, and $\gamma = G_T/A_j$ is the tunnel conductance
per unit area.
The space- and time-dependent matrix $Q({\bf r},\tau,\tau')$
describing electron dynamics in the metal is the
$4\times 4$ matrix in the direct product of the spin space
(subscripts $\alpha$, $\beta \dots$, and Pauli matrices $\sigma$)
and the particle-hole (PH) space (Pauli matrices $\tau$).
In general, $Q$ should also act in the replica space
with the number of relicas $N_r\to0$. However,
for the sake of perturbative calculations which do not
involve closed loops of diffusive modes we can safely set
$N_r=1$ thus omitting the redundant replica space.
The $Q$-matrix obeys the constraint $Q^2 = \openone$
and the symmetry condition $Q = \tau_2 Q^T \tau_2$.
The usual Green functions of disordered metal
correspond to the stationary uniform saddle-point $\Lambda$ of the
action $S_D$ [written in the energy representation, with
$E_m = \pi T (2m+1)$]:
\be
  \Lambda_{\alpha\beta}(m,n) =
  \delta_{\alpha\beta} \delta_{mn} \sign (E_m) \, \tau_3 .
\label{Lambda}
\ee
Equation (\ref{ST}) contains the superconductive matrix $Q_{Sj}$
of the  $j$-th island:
\be
  Q_{Sj}(\tau) =
  \begin{pmatrix}
    0 & \sigma_2 e^{i\vp_j(\tau)} \\
    \sigma_2 e^{-i\vp_j(\tau)} & 0
  \end{pmatrix} .
\ee

Diffusion modes of the disordered metal are accounted for
by the $Q$-matrix fluctuations near the saddle-point $\Lambda$.
They can be parametrized as
\be
  Q = \Lambda
    \left[
      1 + W + \frac12 W^2 + c_3 W^3 + c_4 W^4 + \dots
    \right] ,
\label{param}
\ee
in terms of the antihermitian matrix $W$ obeying the constraint
$\{\Lambda,W\} = 0$, and $c_4=c_3-1/8$.
In the PH space the matrix $W$ is given by
\be
  W = \begin{pmatrix}
    d & c \\
    -c^\dagger & -d^T
  \end{pmatrix} ,
\label{W}
\ee
with $d=-d^\dagger$ and $c=c^T$ describing diffuson and cooperon modes,
respectively. These matrices acting in the spin and Matsubara spaces
are nonzero only
if $\eps_m \eps_n < 0$ (diffusons) and $\eps_m \eps_n > 0$ (cooperons).
Their bare propagators have the form:
\begin{subequations}
\label{corrs}
\begin{align}
  \corr{d_{\alpha\beta}(m,n,\br) d_{\alpha\beta}^*(m,n,\br')}
  & = \frac{2}{\pi\nu} \, \cD(\br,\br',\eps_m,\eps_n) ,
\\
  \corr{c_{\alpha\beta}(m,n,\br) c_{\alpha\beta}^*(m,n,\br')}
  & = \frac{2}{\pi\nu} \, \cC(\br,\br',\eps_m,\eps_n) ,
\end{align}
\end{subequations}
where
\begin{subequations}
\label{D&C}
\begin{align}
\cD(\br,\br',\eps_m,\eps_n)
& = \theta(-\eps_m\eps_n) \, \cD_0(\br,\br',\eps_m-\eps_n),
\label{D}
\\
\cC(\br,\br',\eps_m,\eps_n)
& = \theta(\eps_m\eps_n) \, \cD_0(\br,\br',\eps_m+\eps_n)
\label{C},
\end{align}
\end{subequations}
and $\cD_0(\br,\br',\omega)$ is the Green function of the diffusion operator:
\be
  \left( -D\nabla^2 + |\omega| \right)
  \cD_0(\br,\br',\omega)
  = \delta(\br-\br')
\label{D0}
\ee
with the boundary condition $\nabla_{\bf n} \cD_0(\br,\br',\omega) = 0$
at the NS interface.

\section{Dynamics of the phase}

Integration over cooperon modes in the Gaussian
approximation yields the action functional that describes
phase dynamics of the array~\cite{FL98,FLS01}. For a single island,
this action is of the form ($\omega_k=2\pi T k$):
\be
  S_A = T \sum_k \left[
  \frac{\omega_k^2 |\vp_k|^2}{4E_C}
  + \frac{|\omega_k| G_A(\omega_k)}{8}
    (e^{i\vp})_k (e^{-i\vp})_{-k}
  \right] ,
\label{SA}
\ee
where $E_C=2e^2/C$ is the bare charging energy, with $C$ being
the total island capacitance, and
the Andreev conductance $G_A(\omega)$ is given by Eq.~(\ref{GA})
(here we neglect the interaction-induced
corrections to $G_A$ studied in Refs.~\cite{FL98,FLS01}).

The action (\ref{SA}) had been studied extensively
starting from the pioneering paper~\cite{Kosterlitz}
(cf.\ Ref.~\cite{BAL} and references therein).
At low enough frequencies, $\omega\ll\Omega_{0}$,
where $\Omega_{0}=G_A(\Omega_{0})E_C$,
only the second term in Eq.~(\ref{SA}) is relevant
and the theory becomes logarithmic provided that
$G_A(\Omega_{0}) \gg 1$. The latter condition which prohibits
the Coulomb blocking of tunneling will be assumed hereafter.

Phase dynamics can be characterized by the imaginary-time phase
autocorrelation function $\Pi_M(\tau)=\corr{e^{i\vp(\tau)-i\vp(0)}}$.
This correlator decays at the time scale $\hbar/\tilde{E}_C$,
where $\tilde{E}_C$ is the renormalized effective charging energy,
which is exponentially small in the considered regime of weak Coulomb
blockade.
For $\omega$-independent $G_A(\omega)$ (corresponding to the 3D situation),
the most detailed of
existing estimates for $\tilde{E}_C$ was found in Ref.~\cite{BAL}
using the two-loop renormalization group (RG)
together with the instanton analysis:
\be
  \tilde{E}_C
  \approx
  \frac{E_C}{3\pi^2} \left( \frac{\pi G_A}{2} \right)^4
  \exp \left( - \frac{\pi G_A}{4} \right) .
\label{EC}
\ee
At $T \gg \tilde{E}_C$ the deviation of the autocorrelation function
$\Pi_M(\tau)$ from 1 can be determined by means of RG;
in the one-loop approximation [valid at $\Pi_M(\tau) \gg 1/G_A$]
the result is~\cite{FL98}:
\be
  \Pi_M(\tau) = 1-\frac{4}{\pi G_A}
\ln\left(\frac{G_AE_C}{2\pi^2\hbar}\tau\right) .
\label{Pi}
\ee

In the 2D case, $G_A(\omega)\propto\ln\omega$ which leads to an
extremely slow ($\ln\ln\tau$) correction to $\Pi_M(\tau)$ and,
hence, to negligibly small $\tilde{E}_C$~\cite{FL98}.
To find  $\tilde{E}_C$ one then should take into account that
the simple formula (\ref{GA}) is modified in the lowest-frequency limit
due to i) Cooper-channel repulsion in the normal metal, and ii)
breakdown of the lowest-order tunneling approximation, both these
effects were considered in~\cite{FLS01}. Below in this paper we
assume (for the 2D case) that temperatures are not too low and
approximation (\ref{GA}) is valid.

\section{Phase transition}

\subsection{Thermal transition}

The temperature $T_c(n_i)$ of the thermal superconducting transition
is determined by the mean-field relation~\cite{FL98}
\be
  T_c={\cal J}(T_c)/2, \qquad {\cal J}(T)=\sum_{i} E_J(r_i,T) ,
\label{MF-Tc}
\ee
where $E_J(r,T)$ is the ($T$-dependent) energy of proximity-induced
Josephson coupling between two SC islands at the distance $r$
in $d$ dimensions:
\be
  E_J(r,T)
  = \frac{G_T^2}{8\pi\nu \xi^2_T (2r)^{d-2}}
    \sum_{n=0}^\infty P_d \left( \frac{r}{\xi_T} \sqrt{2n+1} \right) ,
\label{EJ}
\ee
where $\xi_T = \sqrt{\hbar D/2\pi T}$ is the thermal length,
and we denoted $P_3(x)=\exp(-x)$ and $P_2(x)=K_0(x)$.
Equation (\ref{MF-Tc}) is valid if the number of relevant terms
in the sum for ${\cal J}(T_c)$ is large, otherwise the transition
is not of the mean-field type, but Eq.~(\ref{MF-Tc}) can still serve
as an estimate for $T_c$.

The nature of the transition in $d$ dimensions is determined by the parameter
$\delta_d$:
\begin{subequations}
\label{deltad}
\begin{align}
  & \delta_3
  = \frac{G_T^2}{8\nu\hbar D b}
  = \frac{3\pi^2 G_T^2}{4(k_Fl)(k_Fb)}
  = \frac{3 \Gamma^2(k_Fa)^4}{4(k_Fl)(k_Fb)} ,
\label{delta3}
\\
  & \delta_2 = \frac{G_T^2}{8\nu\hbar D} = \frac{G_T^2}{4g} ,
  \label{delta2}
\end{align}
\end{subequations}
which is an estimate for $E_J(b,T)/T$ at $T=\hbar D/2\pi b^2$,
and $b=n_i^{-1/d}$ is the typical distance between the islands.
In Eq.~(\ref{delta3}) we expressed $G_T = \Gamma k_F^2A_j/4\pi^2$
through the characteristic transmission coefficient $\Gamma\ll1$
of the S-I-N tunnel barrier.

In three dimensions the parameter $\delta_3$ can be arbitrary
compared to 1. However, in two dimensions the parameter
$\delta_2$ is bounded from below by the requirement of weak
Coulomb blockade: $G_A(\Omega_0) = (\delta_2/\pi) \ln(l/d_I) \gg 1$,
where we estimated the island's capacity as $C\sim a^2/d_I$,
with $d_I$ being the width of the insulating barrier.
This condition requires $\delta_2\gg1$. Otherwise the transition
is driven by quantum fluctuations and occurs at $E_C\sim {\cal J}$.

If $\delta_d\ll1$ then $T_c\ll D/2\pi b^2$, the Josephson coupling
is long-range and the mean-field equation (\ref{MF-Tc}) gives for
the  transition temperature:
\be
\label{Tc1}
  T_c = \frac{G_T^2 n_i}{16\nu} \ln \frac{1}{\delta_d}
  = \frac{\hbar D}{2\pi b^2} \: \pi\delta_d \ln \frac{1}{\delta_d},
\qquad
  \delta_d\ll1 .
\ee

If $\delta_d\gg1$ then $T_c\gg \hbar D/2\pi b^2$ and the Josephson coupling
is short-range. The transition temperature can be estimated as
\be
\label{Tc2}
  T_c = \frac{\hbar D n_i^{2/d}}{2\pi} \ln^2 \delta_d
  = \frac{\hbar D}{2\pi b^2} \: \ln^2 \delta_d,
\qquad
  \delta_d\gg1 .
\ee

\subsection{Quantum transition}

Quantum transition can be described within the lowest tunneling
approximation only in three dimensions (cf.~\cite{FLS01} for discussion of
quantum phase transition in a more complicated 2D case).
Upon decreasing $n_i$, the
transition temperature defined by Eq.~(\ref{Tc1}) lowers eventually
below $\tilde{E}_C$, then quantum fluctuations should be taken into account.
At some critical concentration $n_c$  the temperature of the superconductive
transition vanishes, marking the point of a quantum phase transition.
The point of the quantum
transition is determined by the equation similar to (\ref{MF-Tc}):
$\tilde{E}_C = {\cal J}(0)$ (cf.~\cite{FL98} for more details).
However, the zero-temperature value of the integrated Josephson
proximity coupling ${\cal J}(0)$ cannot be determined by the simple
formula (\ref{EJ}) due to logarithmic divergency of the resulting
expression. This divergency is cured by the account of the
Cooper-channel repulsion constant in the metal $\lambda_n$~\cite{ALO68}
leading to ${\cal J}(0) = G_T^2n_i/16\nu\lambda_n$.
As a result, the critical concentration $n_c$ is found to be
\be
  n_c = \frac{16\pi\nu \lambda_n\tilde{E}_C}{G_T^2} ,
\label{nc}
\ee
where $\tilde{E}_C$ is defined in Eq.~(\ref{EC}).

\section{Cooperon self-energy}

\begin{figure}
\epsfxsize=75mm
\centerline{\epsfbox{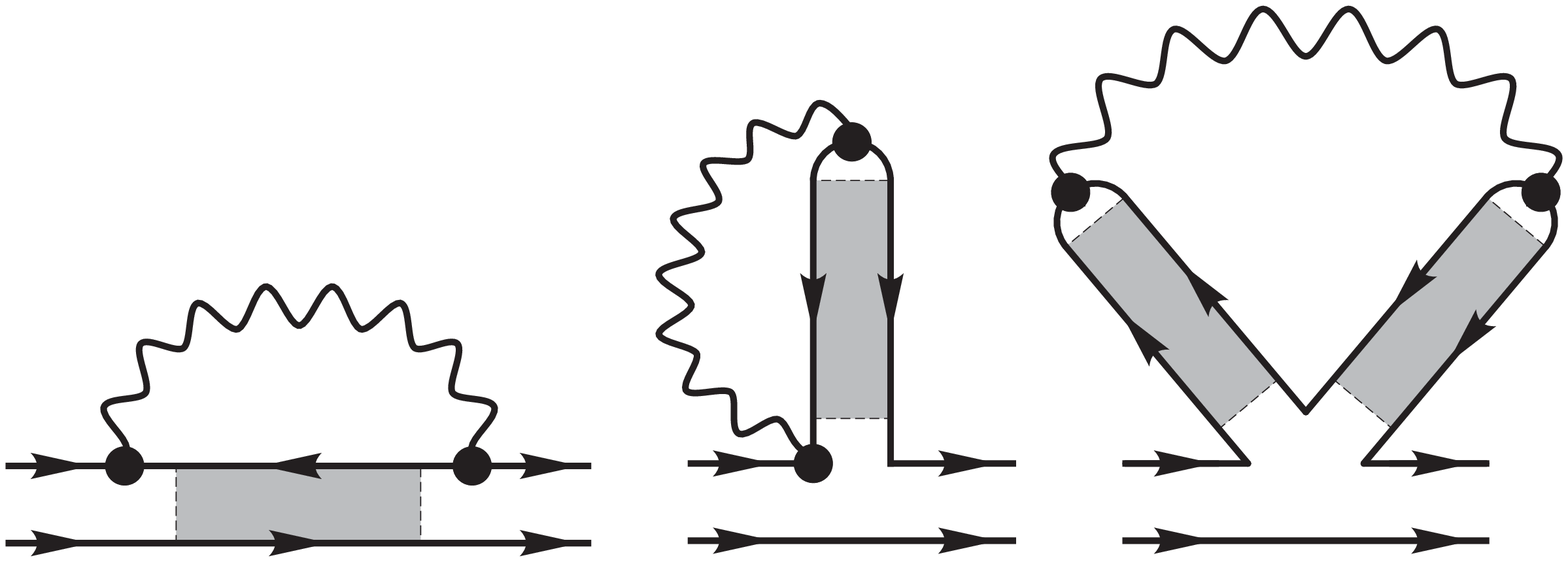}}
\caption{Diagrams for the cooperon self-energy
in the second order over $G_T$.
Shadowed blocks are cooperons and diffusons,
dots denote Andreev reflections from the dot,
and wavy lines stand for the phase correlation
function $\Pi(\omega_k)$.}
\label{F:diagrams}
\end{figure}

In the presence of SC islands, cooperon modes are no longer gapless.
To obtain the cooperon self-energy due to Andreev reflection
we calculate the correction
to the action in the lowest tunnel approximation:
\be
  \delta S =
- \frac{\corr{S_T^{(2)} S_T^{(2)}}}2
- \corr{S_T^{(3)} S_T^{(1)}}
+ \frac{\corr{S_D^{(4)} S_T^{(1)} S_T^{(1)}}}2
,
\label{deltaS}
\ee
where the vertices $S_D^{(l)}$ and $S_T^{(l)}$ come from expansion
of the actions~(\ref{SD}) and (\ref{ST}), respectively,
to the order $W^l$.
The second order in $G_T$ approximation (\ref{deltaS}) is valid provided
that $G_T\ll G_D$ \cite{Nazarov94,SFL01}.
The corresponding diagrams are shown in Fig.~\ref{F:diagrams}.
Their sum is independent of the certain form of the
parametrization (\ref{param}).
Averaging in Eq.~(\ref{deltaS}) goes over phase $\vp_j(\tau)$ dynamics
and bare diffusive modes (\ref{corrs}).
It is important that at $T\gg T_c$ the phases on different islands
are uncorrelated with each other.
Upon averaging, one obtains the cooperon part of the induced action
(\ref{deltaS}), which in the long-wavelength limit takes the form:
\be
  \delta S_C = - \frac{\pi\nu}{4} \,
    T^2 \sum_{mn} \int d\br \:
    \Sigma_{mn} \left|c_{\alpha\beta}(m,n,\br)\right|^2 ,
\label{deltaSC}
\ee
where $c$ is the cooperon part of the matrix $W$, Eq.~(\ref{W}), and
\begin{align}
  \Sigma_{mn} = {} & \frac{n_i G_T^2}{16\nu^2} \,
  T \sum_{k}
  \int \frac{dA\, dA'}{A^2} \,
  [ \cD(\br,\br';m,n-k)
\nonumber \\ {}
  & - \cC(\br,\br';m,m-k) ] \, \Pi_M(k)
  + \{m\leftrightarrow-n\} ,
\end{align}
and $\Pi_M(k)$ is an imaginary-frequency version of the
autocorrelation function $\Pi_M(\tau)$.
Equation~(\ref{deltaSC}) is valid provided that the cooperon wave
vector $q$ is smaller than the inverse separation between
the islands, $q\ll n_i^{1/d}$, which allows to pass from the discrete
sum over the islands to the uniform integration over $\br$:
$\sum_j \to n_i \int d\br$.
The self-energy $\Sigma_{mn}$ determines the low-$q$ behavior
of the cooperon: $\cC(\bq,m,n)=(Dq^2+|\eps_m+\eps_n|-\Sigma_{mn})^{-1}$.

Integrating diffusive modes over the area $A$ of the contacts
yields the normal-metal resistance $G_D^{-1}$ which combines
with $G_T^2$ into the Andreev conductance.
After simple algebra we obtain for $\eps_m, \eps_n>0$:
\be
  \Sigma_{mn}
  =
  - \frac{n_iG_A}{8\nu}
  \: T
  \left[
    \sum_{k=-m}^m \Pi_M(k)
  + \sum_{k=-n}^n \Pi_M(k)
  \right] ,
\label{Sigmam}
\ee
which is written for the case of $\omega$-independent $G_A$.
Analytic continuation of Eq.~(\ref{Sigmam}) from $\eps_m, \eps_n>0$ to real
frequencies, $i\eps_m \to \eps+i0$, $i\eps_n \to \eps'+i0$,
yields the cooperon self-energy
\begin{align}
  \Sigma(\eps,\eps')
  = &
  - \frac{n_iG_A}{8\nu}
    i \int_{-\infty}^\infty \frac{d\Omega}{2\pi}
    \left\{
      \Pi^K(\Omega)
\right.
\nonumber
\\ {}
  &
\left.
    {} + \Pi^R(\Omega) F(\eps-\Omega)
      + \Pi^A(\Omega) F(\eps'+\Omega)
    \right\}
\label{Sigmar}
\end{align}
in terms of the Keldysh, retarded and advanced components
$\Pi^{K,R,A}(\Omega)$ of the phase correlation function,
and $F(\Omega)=\tanh(\Omega/2T)$.

To study the quantum  corrections to conductivity at zero frequency
we set $\eps'=-\eps$ leading to the cooperon decay rate
$\gamma(\eps) = - \hbar^{-1}\Sigma(\eps,-\eps)$:
\be
  \gamma(\eps)
  =
  \frac{n_iG_A}{4\hbar\nu}
     \int_{-\infty}^\infty \frac{d\Omega}{2\pi}
    \Pi(\Omega)
    \left\{
      1 - F(\Omega)F(\Omega-\eps)
    \right\} ,
\label{gamma-f}
\ee
where we used the identity
$\Pi^K(\Omega) = -2i \Pi(\Omega)$, where $\Pi(\Omega)$ is the Fourier-transform
of the real-time symmetrized autocorrelation function of the island's order
parameter $\Pi(t) = \corr{\cos[\vp(t)-\vp(0)]}$.
Another useful representation for $\gamma(\eps)$ follows
from Eq.~(\ref{gamma-f}) by means of the inverse Fourier
transformation into the time domain:
\be
  \gamma(\eps) =
   \frac{n_i G_A}{2\hbar\nu} T \coth\frac{\eps}{2T}
  \int_0^\infty
  \Pi(t)
  \frac{\sin \frac{\eps t}{\hbar}}{\sinh\frac{\pi Tt}{\hbar}} dt .
\label{gamma}
\ee

It is interesting to note that the functional form of Eq.~(\ref{gamma})
coincides exactly with the expression for the tunneling density of states
in the presence of the Coulomb zero-bias anomaly, cf.\ Eq.~(58) of Ref.~\cite{KA}.
In the present case the island's phase $\vp(t)$ plays the role
of the Coulomb-induced phase $K(t)$ introduced in~\cite{KA},
whose fluctuations give rise to the zero-bias anomaly.
Then expression (\ref{gamma}) can be rationalized with simple physical
interpretation: ``superconductive'' contribution to the cooperon decay rate
is just the average rate of Andreev processes which occur in the
system. Indeed, quantum correction to conductivity comes from interference
between different trajectories of the same electron; Andreev reflection
transforms this electron into a hole, therefore destroying
further interference.

\section{Dephasing time}

Now we start to analyze the consequences of the result (\ref{gamma}).
To evaluate the islands' contribution into the dephasing rate,
we need $\gamma(\epsilon \approx T)$.  Behavior of $\Pi(t \sim \hbar/T)$
is governed by the relation between temperature $T$ and the effective
charging energy $\tilde{E}_C$ of SC islands.

\subsection{3D case}

\subsubsection{Moderately high temperatures, $T\geq \tilde E_C$}

At  $T \geq \tilde{E}_C$, the integral
in Eq.~(\ref{gamma}) converges at $t\sim \hbar/T$
where $\Pi(t)$ is given by Eq.~(\ref{Pi}).
As a result, $\gamma(\eps)$ is nearly energy-independent at $\eps\sim T$
and can be identified with the dephasing rate
leading to the 3D result (\ref{result}).
The latter is valid as long as the expression in the brackets
is large compared to unity.

Assuming that $\tau_\varphi^{(0)}(T)$ is given by Eq.~(\ref{tauAAK}),
we can estimate the upper boundary $T_*^{3D}(n_i)$ of the temperature range
where $1/\tau_\vp^A$ is the main contribution to the dephasing rate.
Using the 3D expression for $G_A$ one finds
\be
  T^{3D}_*(n_i) \sim \frac{\hbar}{\tau} \, x_i^{2/3} (\Gamma k_Fl)^{4/3} ,
\label{T3d}
\ee
where $x_i = (4\pi/3) a^3 n_i$ is the volume fraction of the superconductive
material in the matrix. From the low-temperature side
applicability of the 3D result (\ref{result})
is limited by the thermal transition temperature $T_c^{3D}(n_i)$.
Thus the relative width of the temperature window where
Andreev reflection from the SC islands is the leading mechanism of dephasing
is given by the ratio
\be
  \frac{T^{3D}_*(n_i)}{T^{3D}_c(n_i)}
  \approx
  \begin{cases}
  \displaystyle
    \frac{500 \, G_A^{3/2}}{\ln^2(n_i/n_0)} ,
    & n_i \gg n_0 ,
  \\[10pt]
  \displaystyle
  \frac{50 \, G_A^{3/2} (n_0/n_i)^{1/3}}{\ln(n_0/n_i)} ,
    & n_i \ll n_0 ,
  \end{cases}
\label{T/T-3D}
\ee
where we used Eqs.~(\ref{delta3}), (\ref{Tc1}) and (\ref{Tc2}),
and defined $n_0=(8\nu\hbar D/G_T^2)^3$ such that $\delta_3=(n_i/n_0)^{1/3}$.
Large factors in Eq.~(\ref{T/T-3D}) result partly from the large factor
in Eq.~(\ref{T3d}) hidden there in $x_i$ and $\Gamma$, and partly
from writing $\ln\delta_3=(1/3)\ln(n_i/n_0)$.
We see that the condition $G_A\gg1$ guarantees the existence of the broad
temperature range where the dephasing time is nearly temperature independent
and given by $\tau_\vp^A$.

\subsubsection{Lowest temperatures, $T\ll \tilde E_C$}

The region of very low temperatures, $T \ll \tilde{E}_C$,
can be traced only at very small concentration of the island,
$n_i<n_c$ [cf.\ Eq.~(\ref{nc})], where superconductivity
is absent even at $T=0$ due to quantum fluctuations,
Here the integral (\ref{gamma}) converges at
$t \sim \hbar/\tilde{E}_C$ and can be approximated as
\be
  \gamma(\eps) = \frac{n_iG_A}{2\pi\nu}
  \, \eps \coth\frac{\eps}{2T} \int_0^\infty\Pi(t) dt .
\ee
The above integral is of the order of $\tilde{E}_C^{-1}$.
Then the Andreev-reflection contribution to the dephasing rate
can be estimated as
\be
  \frac{1}{\tau_\varphi^{\text{A}}(T)}
  \sim
  \frac{n_i}{2\pi\hbar\nu} \, \frac{T}{\tilde{E}_C} .
\label{lowT}
\ee
Since $1/\tau_\varphi^A$ scales $\propto T$ it always dominates
the standard 3D result (\ref{tauAAK}) at very low temperatures.
However, the crossover temperature, where
$\tau_\varphi^A = \tau_\varphi^{(0)}$,
scales as $n_i^2$ and can be extremely low for small concentration
of the islands.

\subsection{2D case}

As explained above, staying within the lowest tunneling approximation
we can explore only the region of relatively high temperatures,
$(G_T/4\pi g)\ln(\hbar D/a^2T)\ll1$, where fluctuations are thermal.
Substituting $\Pi(t)$ by 1 in Eq.~(\ref{gamma}) we come to the
result (\ref{result}) for the 2D case. Here, contrary to the 3D case
one can neglect the one-loop fluctuation correction $\propto\ln\ln T$
compared to the bare $\ln T$ dependence of $G_A$.

Comparing with Eq.~(\ref{tauAAK}) one finds that the ``superconductive''
contribution to dephasing is dominant at $T \leq T_*^{2D}(n_i)$, where
\be
  T_*^{2D}(n_i) = \pi \hbar D n_i\frac{G_A(T_*^{2D})}{\ln(\pi g)} .
\label{T2D}
\ee
The relative width of this window is then estimated by the ratio
\be
  \frac{T^{2D}_*(n_i)}{T^{2D}_c(n_i)}
  \approx
    \frac{20 \, G_A(T_*)}{\ln(\pi g) \ln^2(G_T^2/4g)}
\ee
and is large since $G_A \gg 1$.

\section{Magnetoresistance}

Experimentally, $\tau_\varphi$ is determined from
the magnetoresistance data.
For 2D systems, the low-field magnetoresistance is governed by
the weak localization (WL) and Maki-Tompson (MT) corrections
which have the same dependence on the magnetic field~\cite{Larkin80}:
\be
  \frac{\Delta R(H)}{R^2} =
  - \frac{e^2}{2\pi^2\hbar} [\alpha - \beta(T)]
  \, Y \left( \frac{4DeH\tau_\vp}{\hbar c} \right),
\label{MR}
\ee
with $Y(x)=\ln(x)+\psi(1/2+1/x)$.
Here $\alpha=1$ ($-1/2$) is the WL contribution in the limit of
weak (strong) spin-orbit interaction, while the MT contribution
is characterized by the function $\beta(T)$ expressed
through the Cooper channel interaction amplitude
$\Gamma(\omega_k)$~\cite{Larkin80}:
\be
  \beta(T)
  = \frac{\pi^2}{4} \sum_m (-1)^m \Gamma(\omega_m)
    - \sum_{n\geq0} \Gamma''(\omega_{2n+1}) .
\label{beta}
\ee
In a uniform system far above $T_c$, $\beta(T)\sim1/\ln^2(T/T_c)$
indicating that the MT contribution is smaller but in general comparable
to the WL contribution.

For our system, effective attraction in the Cooper channel
emerges as a result of Andreev reflection from the SC islands.
Formally, integration over the phases $\vp_j(\tau)$ of the islands
generates the standard Cooper interaction term in the action
with $\Gamma(\omega_k)=(n_iG_T^2/16\nu)\Pi(\omega_k)$,
where we made use of the fact that the phases of different
islands are uncorrelated at $T \gg T_c$ and performed
spacial average justified by the inequality $L_\vp\gg b$.
In the temperature range considered, $\Pi(\tau)$ is nearly
constant on the time interval $\tau\in[0,1/T]$, so one can use
the static approximation $\Pi(\omega_k)=\delta_{k,0}/T$.
Substituting into Eq.~(\ref{beta}) we obtain for $T \gg T_c$
\be
  \beta(T) = \frac{\pi^2}{64} \frac{n_iG_T^2}{\nu T} .
\label{beta1}
\ee
Comparing with the estimate (\ref{T2D}) one finds that
$\beta(T) \gg 1$ at $T \ll T_*$, that is magnetoresistance is mainly due to
the MT term and thus is {\em positive} irrespectively of the strength of
the spin-orbit scattering.

Another relevant contribution to magnetoresistance is the Aslamazov-Larkin
(AL) correction. In the range $T \gg T_c$, using the condition
$\delta_2 \gg 1$, one can estimate $\Delta g_{\text{AL}}  \le \hbar n_i D/T$.
Comparing with $\Delta g_{\text{MT}}$ following from
Eqs.~(\ref{MR}) and (\ref{beta1}),
one finds $\Delta g_{\text{\text{AL}}}/\Delta g_{\text{MT}} \sim 1/\delta_2 \ll 1$.
 Moreover, the
 relevant scale of magnetic field $B_{\text{AL}}$ for the AL contribution to
 magnetoresistance is $B_{\text{AL}} \sim cT/eD$,
i.e., it is by factor
$T\tau_\varphi/\hbar \gg 1$ larger than the corresponding WL scale
$B_{\text{WL}} \sim \hbar c/eD\tau_\varphi $.
Thus AL correction to magnetoresistance is much smaller
than quantum (WL and MT) corrections, and $\tau_\varphi$ can be
extracted from the standard low-field magnetoresistance measurements.

We believe the same conclusion to be valid in the 3D case.
Here, however, the MT correction can be either large or small
compared to the WL correction, depending on temperature and
other parameters of the problem.

\section{Discussion}

We demonstrate that small concentration of superconductive
islands can enhance considerably the low-temperature dephasing rate
in disordered bulk and thin-film metals as seen via the low-field
magnetoresistance. In 2D the dominant quantum correction to
magnetoresistance is the Maki-Tompson one, thus magnetoresistance is
positive in the region of interest. Throughout the whole range where our
results are valid, $T\tau_\varphi/\hbar \gg 1$, which allows to neglect
the energy dependence of the cooperon decay rate (\ref{gamma}).
This is why magnetoresistance follows the standard formula (\ref{MR})
derived for uniform metal films.

It was implicitly assumed while deriving Eq.~(\ref{gamma})
that $L_\vp = \sqrt{D\tau_\varphi}$ is much longer than inter-island
separation $b$.
Using Eq.~(\ref{result}) one finds that
in 2D case for this condition to be fulfilled the tunnel-limit inequality
$G_T/G_D \ll 1$ is required; for 3D case the condition $L_\vp \gg b$  is
less restrictive.
We expect that in the 2D case with SC islands strongly ($G_T\gg G_D$)
coupled to the film~\cite{FLS01} the ``Andreev'' contribution
to the dephasing rate at moderate temperatures
can be estimated analogously to Eq.~(\ref{result}), with the proper
expression $G_A\approx G_D$ for the Andreev conductance,
leading to $1/\tau_\vp^A \sim n_i D/\ln(\xi_T/a)$.
Although we considered temperatures much below the intrinsic
transition temperature of SC islands $T_{c0}$, our approach can be
adapted for $T \sim T_{c0}$.

We note in passing that inhomogeneous in space
superconductive gap function is known to affect the BSC peak in
the density of states in a way very similar to that of magnetic
impurities~\cite{LO71}.
The present results show that analogy between inhomogeneous
superconductivity and magnetic impurities extends to dephasing as well.
The influence of the same
dephasing mechanism  upon other phase-coherent phenomena (e.g., mesoscopic
fluctuations and persistent currents) remains to be studied.

The authors are grateful to T. Baturina, H. Bouchiat, A. Kamenev,
and V. Kravtsov for helpful discussions.
This work was supported by the RFBR grant No.~01-02-17759,
the Russian Ministry of Science and
Russian Academy of Sciences (M.A.S. and M.V.F.), the Dynasty
Foundation, the ICFPM (M.A.S.), and NSF grant No.~01-20702 (A.I.L).

\end{document}